\documentclass[]{article}

\usepackage[margin=3cm]{geometry}
\usepackage{authblk}
\usepackage{graphicx}
\graphicspath{{./figs/}}
\usepackage{amsmath}
\usepackage{amssymb}
\usepackage{cite}
\usepackage{xcolor}
\usepackage[caption=false]{subfig}
\usepackage{caption}
\usepackage{hyperref}
\hypersetup{
	colorlinks = true,
	urlcolor   = blue,
	citecolor  = blue,
	linkcolor = blue
}

\title{Clustering through pair interactions in swimming zooplankton}
\author[1]{Ron Shnapp} 
\author[2]{Fran\c{c}ois-Ga\"{e}l Michalec}
\author[1,3]{Markus Holzner}
\date{}

\affil[1]{Swiss Federal Institute for Forest, Snow and Landscape Research, WSL, Zürcherstrasse 111, 8903 Birmensdorf, Switzerland}

\affil[2]{Univ. Lille, CNRS, Univ. Littoral Côte d'Opale, UMR 8187 - LOG - Laboratoire d'Océanologie et de Géosciences, Station Marine de Wimereux, Université de Lille, France}

\affil[3]{Swiss Federal Institute for Aquatic Science and Technology, Ueberlandstrasse 133, 8600 Dübendorf, Switzerland}

\newcommand{\av}[1]{\langle {#1} \rangle}
\newcommand{\volume}{\makebox[1pt][l]{$-$}V}

\begin{document}

\maketitle

\begin{abstract}
	\noindent This work focuses on the formation of mating aggregates in zooplankton. In particular, sexual encounters are behaviourally supported by males actively swimming in search for females, and approaching them for mating once they are found. While the random search leads to a diffusive flux of individuals, the approaching for encounter supports attraction. Thus, we ask whether these competing mechanisms of diffusion and attraction can support aggregation and lead to the formation of mating clusters. To answer our question we formulate a model in which particles performing random walks can briefly make contact with other particles if they are found within a particular distance from each other. Our analysis shows that this model supports clustering in a way analogous to the process of colloid aggregation. Following that, we analyze a dataset of 3D trajectories of swimming copepods and show that the results compare well with our model. These results support the hypothesis that pair-interactions promote mating aggregates in zooplankton and are sufficient to overcome the diffusive nature of their mate searching behavior. Our results are useful for understanding small-scale clustering of zooplankton, which is crucial for predicting encounter rates and reproduction rates in the ocean.
\end{abstract}

\section*{Introduction}

Zooplankton are no longer considered passive drifters at the mercy of ambient currents~\cite{Guasto2012}. Indeed, zooplankton are known to actively swim to search for food~\cite{VanDuren1995, VanDuren1996, Currie2006}, escape from predators~\cite{Kirboe2014, Yen2015}, migrate to favorable environments (e.g., diel vertical migration, DVM)~\cite{Bjoernsen1991, Maar2003}, and find mating partners~\cite{Yen1998, Kioerboe2005, Visser2007, Kiorboe2018}. Therefore, the spatial distribution of zooplankton depends not only on physical processes (e.g., \cite{Levin1976, Gower1980, Abraham1998, Bode1996}), but on behavioral traits as well~\cite{Painter2021, PinelAlloul2007, Strickler1998, Ritz1994} -  a critical feature in marine ecology.

As Folt and Burns~\cite{Folt1999} discussed, the four behaviors specified above can lead to patchiness in the plankton spatial distribution at different scales. The vertical migration of certain zooplankton species due to DVM or escaping high levels of near-surface turbulence generates a vertical dimension of patchiness in zooplankton distributions at the $\mathcal{O}(10-100)$ meters scale. Predators can promote patchiness of spatial distributions directly by removing individuals, and indirectly by eliciting avoidance behaviour that might trigger DVM and alter small-scale patchiness by regulating the individuals' swimming characteristics. Food searching behaviour can also promote patchy zooplankton distributions at various scales if the food source is itself patchy. Lastly, field observations showed the existence of small-scale monospecific copepod clusters that were associated with mating behavior~\cite{Ratzlaff1974, Young1978, Davis1992}. We shall focus on the last of the four factors - patchiness due to mating behavior.

Let us consider what are the phenomena that can lead to aggregarion of particles at a mechanistic level. Chemotaxis, the attraction of individuals to regions with a higher concentration of certain chemical compounds, is a known driver of aggregarion~\cite{Stocker2012a}; chemotaxis might be supported in zooplankton due to pheromones used for signaling in certain species~\cite{Bagoeien2005, Yen1998}. Flocking can lead to clustering if the motion of individuals is spatially correlated~\cite{Vicsek1995, Giardina2008, Puckett2015}, although there is no evidence that this occurs in plankton. Rehotaxis, the orientation of swimming with respect to the flow velocity, too can lead to aggregation of individuals~\cite{Painter2021}, although it is not clear if it leads to clustering in zooplankton. In this work we shall consider another possible source of aggregation: pair-interactions. In particular, the formation of mating aggregates might be driven by the attraction of males and females that exists in certain species. For example, depending on the species, males can sense females in their environment through either mechanical sensing of hydrodynamical signals generated by swimming females~\cite{Kirboe2014, Michalec2020}, and chemosensing of pheromones~\cite{Poulet1982, Yen1998, Bagoeien2005}. Males will actively search their environment for these cues, and once a signal is acquired, they will try to make contact. Therefore, we ask whether this process of searching and approaching for contact can lead to the observed mating aggregation effect?

There are two aspects that we consider in answering our question. On the one hand, the male's approach for contact does support aggregation simply due to the attraction between individuals. But on the other hand, in their active search, males effectively make random walks to explore their environment, which manifests as a diffusive motion that could make mating clusters less probable. To demonstrate this, let us consider a simple Marlovian random walk in which a particle with position $\vec{x}_i = (x_i,\,y_i,\,z_i)$, moves in discrete steps. All the steps have the same length $S=|\vec{x}_{i+1}-\vec{x}_{i}|$, but their direction is random; this rule can be written as follows:
\begin{equation} \label{eq:RW}
\vec{x}_i = \vec{x}_{i-1} + S\cdot\hat{\theta}
\end{equation}
where $\hat{\theta}$ is a vector of length one and a random direction. For this case, which corresponds to Pearson's original random walk~\cite{Pearson1905}, the mean distance between pairs of particles in an initially close group will increase as the square root of time, $t^{1/2}$; since particles diffuse away from each other, clusters eventually break. This consideration suggests that supporting the persistence of mating clusters requires a balance between attraction and searching. Nevertheless, whether such a balance could be sustained was not examined in the past.

In this work, we explore the balance between searching and attraction using a simplified numerical model and laboratory measurements of calanoid copepods' (\textit{Eurytemora affinis}) trajectories~\cite{Michalec2017, Michalec2020}. Our model reproduces the reduced diffusivity of attracting particles, and, furthermore, it shows that mating clusters can emerge. The model is then verified against experimental laboratory results, showing good agreement between the observed patchiness of the copepods in the model and the experiment. This confirms that the searching-attraction behavior supports mating clusters and is consistent with previous field observations~\cite{Ratzlaff1974, Young1978, Davis1992}.

\section*{Methods}

\subsection*{The pair-interaction model}

We model the motion of individual plankters using an augmented version of eq.~\eqref{eq:RW} which provides a simple framework to investigate the balance between searching and pair-interactions. Thus, in the basic case, namely, excluding pair-interactions, the plankters in our model move in discrete jumps of a fixed size and random direction. Notably, zooplankton swimming is often a complex process, exhibiting fractal trajectories and intermittent jumps~\cite{Schmitt2001, Viswanathan2002, Michalec2015}, however, these features are not necessary to reproduce the detailed balance we are looking for. In addition, different species might exhibit different swimming characteristics, so the more general our model will be, the more we will be able to generalize our conclusions to a broader range of species.

Mate finding in copepods can be divided into two steps~\cite{Michalec2020}. First, males actively search their environment for other individuals. Second, when the male senses an individual in his so-called detection radius, it will approach it and attempt to evaluate whether he should attempt to mate~\cite{Strickler1998}. If this is the case, the male will attempt to catch the individual and mate. Interestingly, cases of males attempting to catch other males were documented as well~\cite{Cheng2008}.

\begin{figure}
	\centering
	\includegraphics[height=6.5cm]{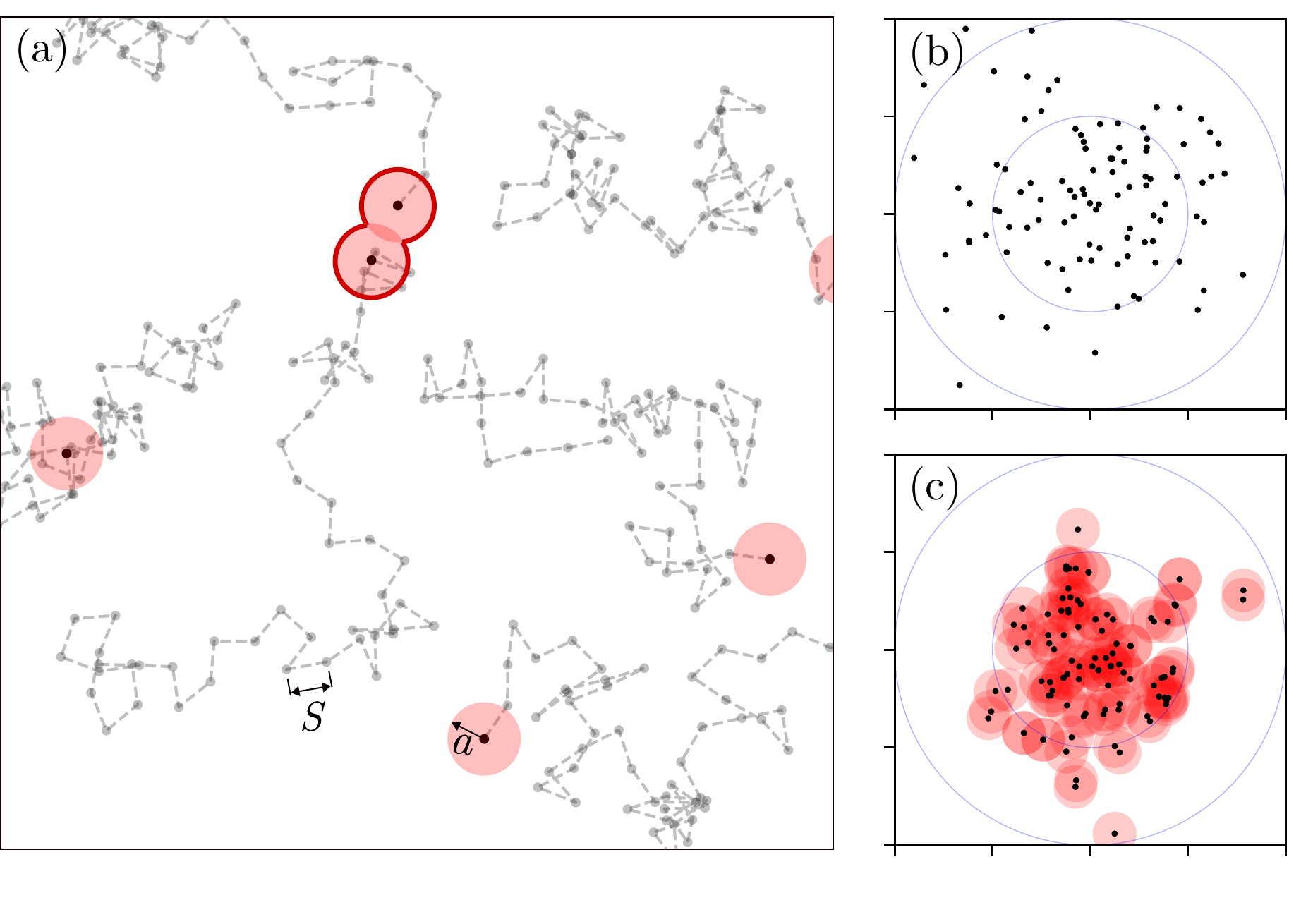}
	\caption{Demonstrations of the pair-interaction model. Black points show the locations of plankters on the $x-y$ plane, and their interaction sphere is shaded in red. (a) Fifty steps long trajectories are shown as dashed lines. Two highlighted plankters are interacting at the last time step. (b) Step number $n=500$ of a simulation with 100 non-intercting plankters ($\rho=0$, and $\mu\rightarrow \infty$). All pplankters start at the origin and they disperse according to eq.~\eqref{eq:model}; two concentric circles with radi $\sqrt{n}\,S$ and $2\sqrt{n}\,S$ denote one and two times the theoretical RMS displacement (eq.~\eqref{eq:msd}). (c) Step number $n=500$ of a simulation with 100 particles, and parameters $\rho=5$ and $\mu=0.5$. The initial conditions and the two concentric circles are the same as in (b). \label{fig:model_ex}}
\end{figure}

Our hypothesis is that the male's behavior of approaching conspecifics constitutes a pair-interaction that can support small-scale clustering. Therefore, to include this mechanism in our model, we add a term to eq.~\eqref{eq:RW}. Overall, our model uses the following discrete evolution equation for the position of plankter $n$: 
\begin{equation} \label{eq:model}
\vec{x}_i^{\,n} = 
\vec{x}_{i-1}^{\,n} +  
\hat{\theta}_i^{\,n} \cdot S +
\sum_{m \neq n} \mathcal{J}_i^{m,n} \cdot \left[\left(\frac{\vec{x}_{i-1}^{\,m} - \vec{x}_{i-1}^{\,n}}{2} \right) - \hat{\theta}_i^{\,n} \cdot S \right] \, ,
\end{equation}
where $S$ is the step size, and $\hat{\theta}_i^{\,n}$ is the random unit vector. The variable $\mathcal{J}_i^{m,n}$ is an index that is equal to zero if plankter $n$ and $m$ do not interact at time step $i-1$, and one if they do. Thus, according to eq.~\eqref{eq:model}, two plankters that interact in time step $i-1$ are moved to occupy the same position at time step $i$. Furthermore, we define that interactions between individuals occur based on two criteria - proximity and a memory kernel. First, two plankters may interact only if the distance between them is smaller than a fixed \textit{interaction radius}, denoted as $a$, and if the distance between them is smaller than the distance to any other plankter. Second, once a plankter undergoes an interaction it cannot interact with any other plankter for a duration of time smaller than a fixed \textit{memory time}, denoted as $M$. Therefore, the interaction index is specified as follows:
\begin{equation} \label{eq:interactions_1}
\mathcal{J}_i^{m,n} = \mathcal{J}_{i, \, \text{distance}}^{m,n} \cdot \mathcal{J}_{i, \, \text{memory}}^{m,n}
\end{equation}
where
\begin{equation} \label{eq:interactions_2}
\mathcal{J}_{i, \, \text{distance}}^{m,n} = 
\begin{cases}
1 \, ,& \qquad \text{if} \quad (d^{\,m,n} < a) \, \text{ and }\, (d^{\,m,n} < d^{\,k,n} \text{ for all } k \not\in \{m, \,n\}) \\
0 \, ,& \qquad \text{otherwise} 
\end{cases}
\end{equation}
with $d^{\,k,n}_i = |\vec{x}_i^{\,k} - \vec{x}_i^{\,n}|$ being the distance between two individuals, and where 
\begin{equation} \label{eq:interactions_3}
\mathcal{J}_{i, \, \text{memory}}^{m,n} = 
\begin{cases}
1 \, ,& \qquad \text{if} \quad (i - i_\mathcal{J}^{m} > M) \, \text{ and }\, (i - i_\mathcal{J}^{n} > M)\\
0 \, ,& \qquad \text{otherwise} 
\end{cases}
\end{equation}
with $i_\mathcal{J}^{n}$ being the largest time step for which $\mathcal{J}_{i_\mathcal{J}^{n}}^{n,k}=1$ for any $k$. An illustration of the pair-interaction model is shown in Fig.~\ref{fig:model_ex}a.

The pair-interaction model, eq.~\eqref{eq:model}--\eqref{eq:interactions_3}, is characterized by three parameters: $a$, $S$ and $M$. Using these fundamental parameters we can form non-dimensional groups that will facilitate our analysis. The first parameter is the dimensionless interaction radius:
\begin{equation}
\rho \equiv \frac{a}{S} \,\, .
\label{eq:rho}
\end{equation}
For $\rho<1$, the steps taken are smaller than the interaction radius so the swept volume of each particle traverses across space continuously. In contrast, for $\rho\gg1$, the swept volume of the particles' interaction radius is non-continuous. For this reason, $\rho$ can be interpreted as the resolution of the particles' search.

To normalize the memory time, $M$, we seek an expression for the typical distance traveled within a period of this duration. Let us first note that the mean translation of particles is zero: $\av{\vec{x}_i - \vec{x}_0}=0$, where $\av{\cdot}$ denotes an ensemble average. In addition, excluding interactions, the root mean squared (RMS) particle translation during $M$ steps is $S_M \equiv \sqrt{M}\,S$ (see Appendix~\ref{App:MSD_proof}). Using this result, we define the following dimensionless parameter:
\begin{equation}
\mu \equiv \frac{S_M}{a} = \sqrt{M} \, \frac{S}{a} = \frac{\sqrt{M}}{\rho} \,\, ,
\label{eq:mu}
\end{equation}
namely, the ratio between the RMS distance traveled during the memory time and the interaction radius.

The importance of the memory time in the pair-interaction model can be appreciated by considering extreme values of $\mu$. For $0<\mu\ll1$, particles that have interacted once will have little chance of "escaping" each other because they do not have sufficient time to leave each other's interaction sphere in the duration of the memory time. Indeed, the limit $\mu\rightarrow 0$ can be thought of as a condensation regime since an infinite-time simulation will result in large patches of particles. In the limit $\mu\rightarrow \infty$, only a very small fraction of the particles will be able to interact at any given time step, meaning that de-facto particles do not interact. A demonstration of this is shown in Fig.~\ref{fig:model_ex}. While the non-interacting particles in subfigure (b) diffuse freely from the origin and increase the separation distance from their neighbors, the interacting particles in subfigure (c) ($\mu=0.5$) diffuse from the origin at a lower rate, and most of the particles have at least one neighbor inside their interaction sphere.

Below, we study the performance of the pair interaction numerically. We used a three-dimensional cubical domain with sides $(0,L)$ and periodic boundary conditions in all calculations. The simulations were performed with 1000 particles, and we maintained a fixed ratio $\frac{a}{L}=(\frac{1}{2\,N_p})\approx 0.0794$, which results in an average concentration of $\frac{1}{2}$ particles per an interaction sphere volume. Furthermore, we let simulations run for a sufficient number of steps to reach a statistical steady state, which is identified by examining time series of the clustering index $I_i$. In particular, we want the particles to have sufficient time to reach and interact with other particles, so the condition $\sqrt{n}\cdot S \gg a$ was maintained ($n$ is the number of time steps in a simulation, see eq.~\eqref{eq:msd}).

\subsection*{The experiment}

We compare the pair-interaction model using a dataset of Calanoid copepod trajectories, \textit{Eurytemora affinis}. The data was captured in a series of laboratory experiments that were previously reported in Refs.~\cite{Michalec2017, Michalec2020}. Here, a mixed-sex population of copepods were freely swimming in still water inside of an aquarium, roughly three liters in volume. The three-dimensional copepod trajectories were obtained through the 3D Particle Tracking Velocimetry (3D-PTV) method using a single camera and an image splitter~\cite{Hoyer2005}. The animals were illuminated from below using a 532 nm laser within a volume of $6\times6\times2$ cm$^3$, and images were captured at a rate of 200 Hz. The measurement lasted 4.5 minutes, with a total of about $10^7$ samples.

The Copepod \textit{Eurytemora affinis} is about 1.0--1.5 mm in size and is a common zooplankton in the low salinity zone of many European and North-American estuaries.~\cite{Kipp2021}. Individuals swim by conducting a series of jumping maneuvers. The jumping frequency and jump lengths vary strongly across individuals and from jump to jump of the same individual. For individuals in our dataset, the jumping distance is about 1 mm on average, although it has a wide tailed distribution that reaches above 8 mm. In addition, the average waiting time between consecutive jumps is around 0.25 seconds, reaching values as low as 0.05 seconds. Notably, previous investigation \cite{Michalec2017} showed that the copepods often conduct a series of short jumps with a correlated direction which leads to an overall larger displacement than the statistics reported here. In addition, the jumping statistics depend on the animals' sex~\cite{Takahashi2005} and on the turbulence conditions in the tank~\cite{Michalec2017}.

The average number of animals in each frame is roughly 200, which corresponds to $\sim2.7$ individuals per centimeter cubed. \textit{Eurytemora affinis} can locate other individuals through mechanical sensing of hydrodynamical signals and through chemosensing of pheromones released by females. Given the high density of copepods in the dataset as compared to the concentration of copepods in the ocean, it seems plausible that most interactions occurred based on mechanical sensing. In addition, based on a similar dataset, Ref.~\cite{Michalec2020} showed that the radial velocity between pairs of \textit{Eurytemora affinis} is negative for individuals sufficiently close to each other, thus estimating their interaction radius for mechanical sensing as roughly 4 mm.

\subsection*{Analysis}

There are two aspects of dispersion that we would like to examine. The first is the effect of pair interactions on the diffusion of particles in the stationary frame of reference, namely, on the typical regions covered by the random walkers in their search. For that, we shall use the theoretical result, that the mean squared displacement of particles (aka dispersion) grows in time according to $\av{\left(\vec{x}_i - \vec{x}_0\right)^2} = i \, S^2$ (see Appendix~\ref{App:MSD_proof}). Using this relation, and in analogy to the Brownian motion, we define an effective diffusivity
\begin{equation}
D_i \equiv \frac{\av{|\vec{x}_i - \vec{x}_0|^2}}{2i} \,\,.
\label{eq:diffusivity}
\end{equation}
For the non-interacting case, the diffusivity does not change with time and it equals $D= \frac{1}{2}S^2$. In our simulations, however, the diffusivity may change in time due to transient effects. We shall compare estimations of $D_i$ for interacting particles and compare them against the theoretical result for non-interacting particles.

The second aspect we shall examine is the ability of pair interactions to lead to clustering. We thus consider $Q_l(k)$: the probability of finding $n=k$ particles inside of a box of side-length $l$. In the case of a random distribution of particles, i.e., particle coordinates are chosen from a uniform random distribution within a fixed volume, $Q_l(k)$ is the Poisson distribution~\cite{Schmidt2017}:   
\begin{equation}
Q_l(k) = \text{Pois}(k,\lambda_l) = \frac{\lambda_l^k}{k!} \exp(-\lambda_l) \qquad ; \qquad 
\lambda_l = \frac{N}{\volume} l^3 \,\, .
\label{eq:Plk}
\end{equation}
Here, $\volume$ is the total volume considered and $N$ is the total number of particles in the tank, so the parameter $\lambda_l$ is equal to the average number of particles in each cell. Using the property eq.~\eqref{eq:Plk}, we can define a so-called clustering index as follows:
\begin{equation}
I \equiv \frac{\av{(k - \lambda_l)^2}}{\lambda_l} \,\, .
\label{eq:clustering_index}
\end{equation}
For a Poisson random variable with parameter $\lambda_l$, both the mean and the variance are equal to $\lambda_l$. Therefore, clustering can be detected for particle distributions where $I>1$, while the opposite could be detected if $I<1$. Note that to allow comparison between different runs, we shall use a fixed value of $l=\frac{1}{10}L$ in our simulations. In addition, the probability that a cluster is of size $k$ can be calculated from $Q_l(k)$ with the normalization:
\begin{equation}
P_l(k) = Q_l(k) / \sum_{k=1} Q_l(k) \qquad \text{for} \qquad k\geq1 \,\,.
\end{equation} 
The distribution $P_l(k)$ will be examined to detect finer details of the particle distribution.


\section*{Results}

\subsection*{Dynamic cluster formation in the pair-interaction model}

\begin{figure}
	\centering
	\subfloat[]{\label{fig:fluct_ci}
		\includegraphics[width=2.8in]{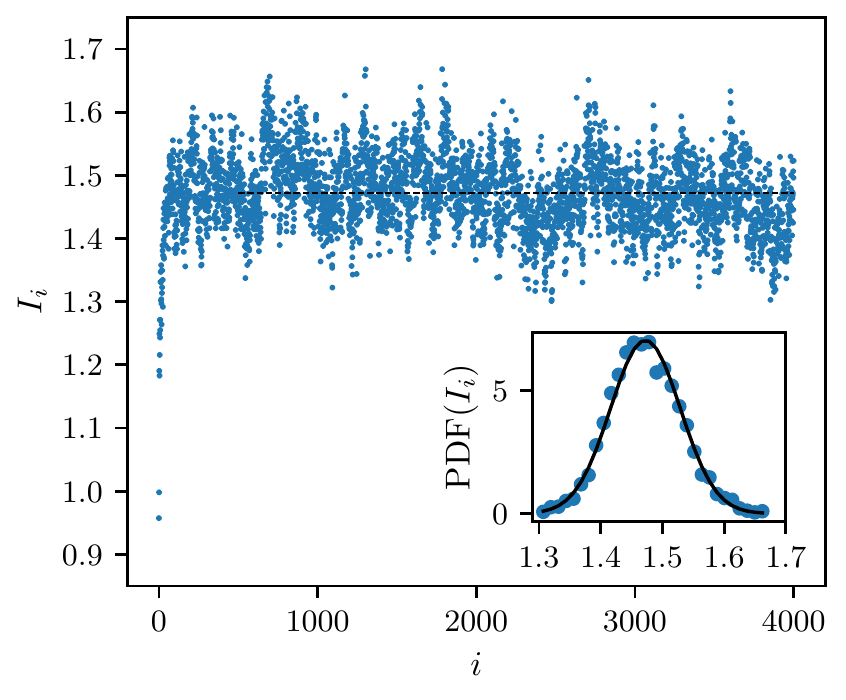}}
	\hfill
	\subfloat[]{\label{fig:dispersion}
		\includegraphics[width=2.8in]{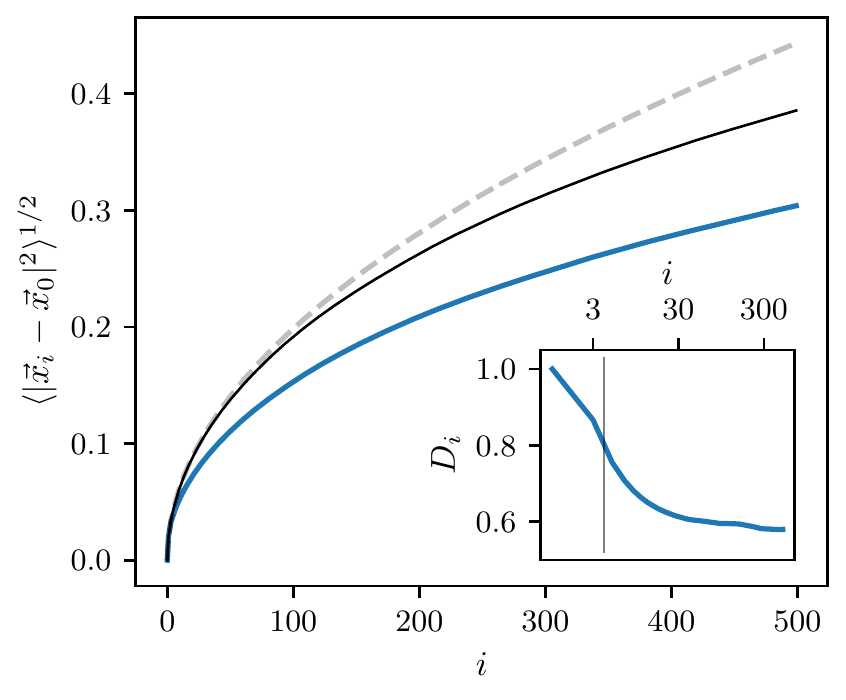}}
	\caption{(a) The clustering index, $I_i$,  is plotted as a function of time step for a simulation with $\rho=4$ and $\mu=0.5$. The mean value of $I_i$, excluding the first 500 samples is shown as a dashed line and the PDF of these values is shown in the inset along with a Gaussian fit. (b) main - dispersion for interacting (blue), non-interacting (black), and the theory for non-interacting particles (dashed line); inset - the ratio between the diffusivity of the interacting and non-interacting cases. \label{fig:basic_characters}}
\end{figure}

Let us demonstrate several basic characteristics of the pair-interaction model using the results of a simulation with parameters $\rho=4$ and $\mu=0.5$. Figure~\ref{fig:fluct_ci} shows the clustering index, $I$, as a function of time, and demonstrates that the clustering index changes with time. All simulations begin with a uniform random distribution of particles in the domain, so at small times, near $i=0$, the clustering index is approximately $I=1$. As time increases, $I_i$ rapidly grows which suggests that particles in the model form clusters. Eventually, for $i\gtrsim500$ in Fig.~\ref{fig:fluct_ci}, the clustering index reaches a quasi-steady regime, where it fluctuates in time around some constant average value. The distribution of the fluctuations around the average, shown in the inset of Fig.~\ref{fig:fluct_ci}, have a distribution very close to the normal distribution, where values of $I_i$ lie between 1.3 and 1.65. The observation of fluctuations in $I_i$ are important, since they suggest there is a dynamical state in the model, in which high and low concentration patches form and break.

In Fig.~\ref{fig:dispersion} we plot the dispersion of the particles in the same simulation. Data is shown for times $i>500$, the range in which quasi-stationarity was observed for $I_i$. In addition, eq.~\eqref{eq:msd} is shown as a grey dashed line, and a thin black line shows the dispersion of particles in a simulation of non-interacting particles ($\rho=0$) with the same step size. In both simulations, the dispersion is slower than what theory predicts in eq.~\eqref{eq:msd}. The slower dispersion for the non-interacting particles occurs due to the finite domain size, which inevitably limits the translation these particles can undergo. In addition to that, the dispersion of the interacting particles is seen to be much slower than that of the non-interacting particles. This is quantified using the ratio between the diffusivities, $D_i$, of the non-interacting and the interacting particles in the inset of Fig.~\ref{fig:dispersion}. At timescales shorter than the memory time ($i=4$, shown as a grey line) the diffusivity of the interacting particles decreases rapidly, and it seems to reach an approximately steady-state only for times longer than the memory time, $i>M=4$. At this long time regime, the diffusivity of the interacting particles is significantly lower, by nearly 50\%, than that of the non-interacting particles. Thus, Fig.~\ref{fig:dispersion} shows that pair-interactions significantly reduce the typical distances traversed by single particles.

\begin{figure}[h]
	\centering
	\subfloat[]{\label{fig:clusters_PDF}
		\includegraphics[width=2.8in]{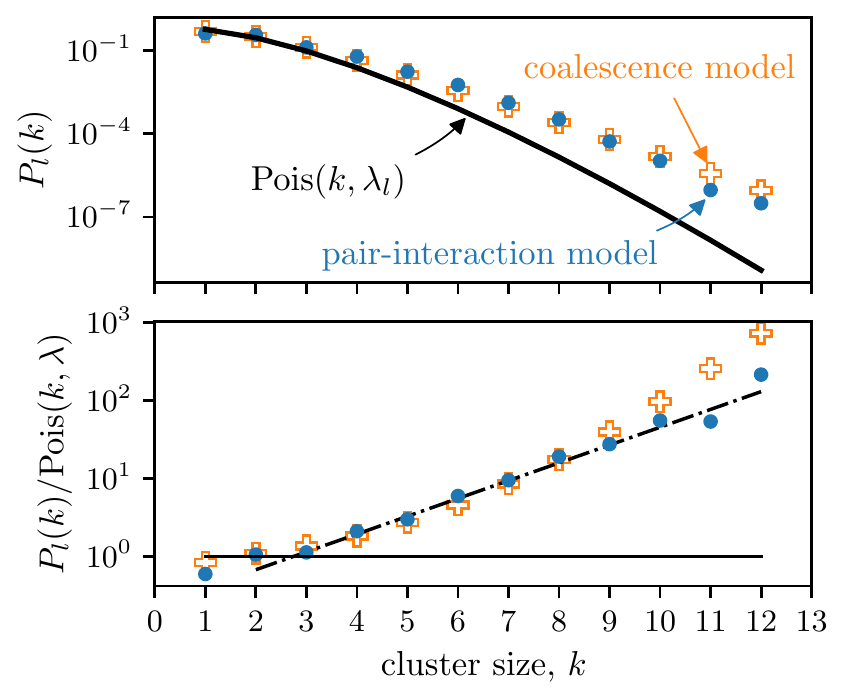}}
	%
	\subfloat[]{\label{fig:cluster_size}
			\includegraphics[width=2.8in]{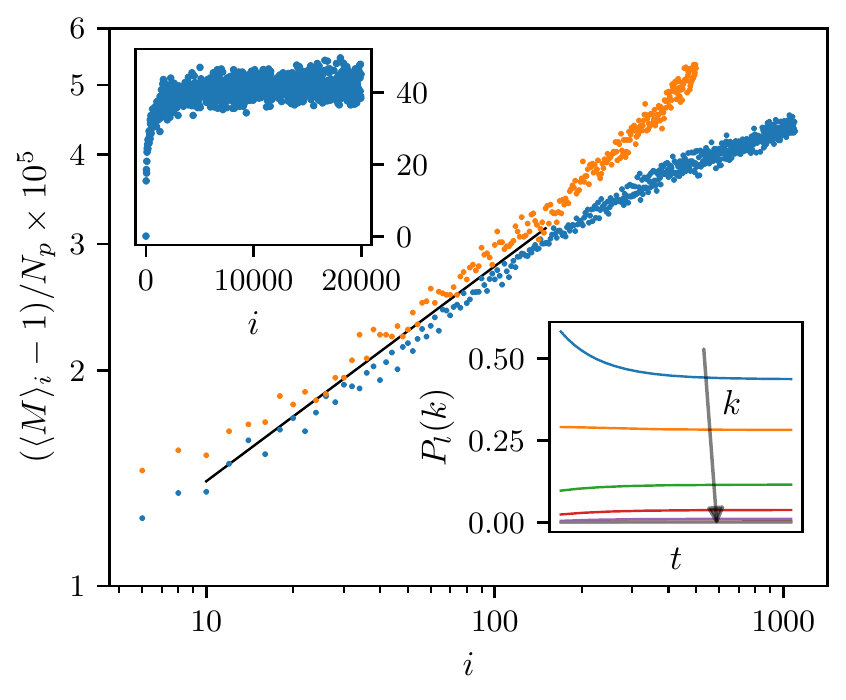}}
	\caption{(a) Top - PDFs of the cluster size $k$ plotted in lin-log scales. Data is shown for the pair-interaction model simulation with $\rho=4$ and $\mu=0.5$, for the clustering model, eq.~\eqref{eq:coagulation_breaking}, and for the Poisson distribution. Bottom - the ratio of the same numerical PDFs and the Poisson distribution; a dot-dashaed line shows an exponential fit to the pair-interaction simulation results, $\exp[\frac{k-2.75}{1.9}]$. (b) Bottom right inset - the evolution of the probability to observe a cluster of size $k$ in the kinetic clustering model, plotted as a function of time for $k=$1, 2, 3$\ldots$ from top to bottom. Main panel - the average number of particles in an aggregate above one, $\av{M}_i-1$, normalized by the total number of particles in the simulation, $N_p$. Data is plotted versus time for two low-density simulations with 10,000 and 3000 particles. A power-law fit to the data with slople of 0.3 is shown as a black line. Top left inset - the normalized average number of particles in a cluster for a long simulation with 1000 particles, showing the approach towards a plateau at the steady-state regime.}
\end{figure}

A key characteristic of the pair-interaction model is the appearance of clusters, regions of high local concentration. The appearance of clusters can be observed in Fig.~\ref{fig:clusters_PDF} through the PDF for having $k$ particles in a cube of side-length $l$, $P_l(k)$. The top panel shows $P_l(k)$ for the interacting particles simulation ($\rho=4$ and $\mu=0.5$ as before) and the Poisson distribution, eq.~\eqref{eq:Plk}, which represents the theoretical homogeneous particle distribution case. Furthermore, the bottom panel of Fig.~\ref{fig:clusters_PDF} shows the ratio between the two PDFs. The figure shows that $P_l(k)$ for the interacting particles is significantly higher than for a homogeneous distribution case. Specifically, the ratio between the PDFs increases roughly exponentially with the cluster size for $k\gtrapprox3$, as shown by the dot-dashed line in the bottom panel. The exponential increase means, for example, that the probability for observing a cluster of $k=12$ particles in the simulation was higher by a factor of about 200 as compared to the homogeneous distribution, and in fact, had clustering not occurred the probability for observing such clusters in the simulation is practically zero. Therefore, Fig.~\ref{fig:clusters_PDF} demonstrates that interactions between pairs of particles cause the formation of clusters with much more than two particles. This is a key feature of the pair-interaction model.

To explain how interactions between pairs of particles lead to high concentration patches, with $k\gg2$, we use a kinetic model that draws on an analogy with the coagulation of colloids in liquid suspension~\cite{Chandrasekhar1943}. We thus denote the concentration of clusters with $k$ particles as $\nu_k$, and use a kinetic model to describe how $\nu_k$ changes with time. There are two processes that govern the kinetics: cluster formation and cluster breakage. For cluster formation, we assume that when two clusters of $i$ and $j$ particles are found at distances smaller than an interaction radius, $R$, they combine to form a new cluster with $k=i+j$ particles. If the clusters are further assumed to move as Brownian particles, the Smoluchowski theory gives the rate of formation of $k$-particle clusters as $4\pi \, R_{ij} \, D_{ij} \nu_i \nu_j$, where $D_{ij} = D(\frac{1}{i} + \frac{1}{j})$ is the relative diffusivity of the two clusters and $R_{ij}$ is their interaction distance~\cite{Chandrasekhar1943}. For the cluster breakage, we assume that a cluster with $k$ particles might break into two clusters of sizes $i$ and $j=k-i$. Then, we use an empirical breaking rate 
\begin{equation}
\beta_{k\rightarrow i,j} = B \cdot k \, \left[ 1 - \text{exp}\left(-\frac{A}{z_{ij}}\right) \right] 
\end{equation}
where $z=(R_{ij}^2/4\,D_{ij}\,M)^{1/2}=(\sqrt{2}\mu)^{-1}$, and $A$ and $B$ are tunning parameters. The breaking rate increases with the number of particles in a cluster and with the memory time, $M$. Thus, following~\cite{Chandrasekhar1943}, and assuming that breaking is a first-order kinetic~\cite{Baebler2008}, we obtain the following equations that describe the time evolution of clusters concentration:
\begin{equation}
\frac{d \, \nu_k}{dt} = 4\pi \left[ 
\frac{1}{2} \sum_{i+j=k}  \nu_i \nu_j D_{ij}\,R_{ij} - \nu_k \sum_{j}^{\infty} \nu_j D_{jk} \,R_{jk} \right] +
\left[ 
\sum_{j=k+1}^{\infty} \nu_{j} \, \beta_{j \rightarrow k,j-k} 
- \frac{1}{2} \sum_{i+j=k} \nu_{k} \, \beta_{k \rightarrow i,j} 
\right]  .
\label{eq:coagulation_breaking}
\end{equation}
Notably, cluster formation and breaking occur at different rates so they may balance each other and reach a steady state.

In Fig.~\ref{fig:clusters_PDF} we plot the PDF that was obtained by numerically solving eq.~\eqref{eq:coagulation_breaking} following fitting of the parameters $A=0.025$ and $B=0.005$ to the PDF $P_l(k)$ obtained with the pair-interaction model. The two PDFs we obtained using the two models are seen to agree reasonably well with each other. Furthermore, the evolution of $\nu_k$ versus time is plotted in the bottom right inset of Fig.~\ref{fig:cluster_size}, showing that similar to the pair interaction model's results, a steady-state regime is achieved in the kinetic model as well. The agreement between the two models and the existence of steady-states suggest that the analogy between the pair-interaction model and the picture of formation-breakage of clusters is valid.

Aggregation dynamics in colloids, depending on the regime of aggregation, are characterized by universal properties~\cite{Lin1989}. In one of the regimes, the diffusion-limited aggregation (DLA), the cluster growth rate is limited by the diffusion of the particles in the medium, which constrains the growth rate of aggregates and leads to a power-law increase of the aggregate's mass with time. To characterize the transient dynamics of the aggregation in the pair interaction model, we performed simulations at higher particle numbers and low particle density with $a/L=0.02$. Then, as an analog to the mean aggregate mass, we calculated the average number of particles in the aggregates, $\av{M}_i$. Thus, we divided the particles' distribution into bins, calculating the average number of particles in each bin over all bins with at least one particle. In the calculations, we used a bin size of $\frac{1}{2}a$, and we did not observe significant changes when varying it. In Fig.~\ref{fig:cluster_size} we plot the normalized mean cluster size as a function of time from two simulations. In the initial transient period in which clusters first form, here $i\lesssim150$, the average cluster size grows algebraically with time, according to the power-law $\av{M}_i-1 \propto i^{\alpha}$, with $\alpha=0.3\pm0.01$. As shown in the top left inset of Fig.~\ref{fig:cluster_size}, at longer times the slope graduly decreases and eventually plateaus when a steady-state regime is obtained. The algebraic growth of the average cluster size completes the analogy between the pair interaction model and colloid aggregation, showing that the particles in the model aggregate according to the DLA dynamics.


The similarity observed between the two models still requires some explaining, since while the basic unit in the pair-interaction model is individual particles, in the kinetic model the basic units of interaction are clusters. To reconcile this difference we recall the reduced diffusivity observed in Fig.~\ref{fig:dispersion}. Considering a group of initially close particles and given that $\mu$ is sufficiently small, repeated interactions between the group's members tend to keep the particles close to each other. It is thus harder for any particle to escape the group (i.e., increasing its distance to the group's center of mass above some multiple of the interaction radius) as the number of particles in the group increases, thus effectively forming a cluster. By repeating this argument for two groups of particles that by random chance happen to be found close to each other, we understand the formation of clusters in the kinetic model through group merger. Lastly, when individual particles do manage to escape the group cluster breakage is obtained.

\begin{figure}
	\centering
	\includegraphics[width=5.6in]{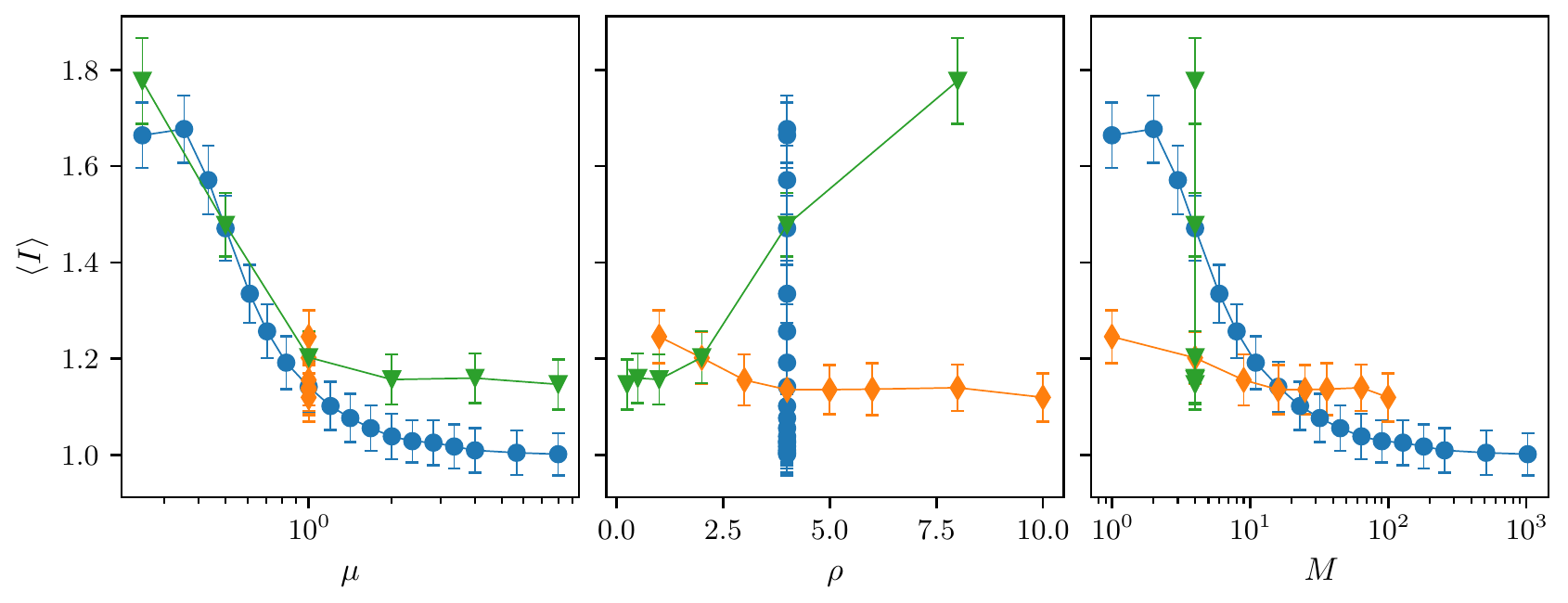}
	\caption{Dependence of the clustering index in the steady-state regime on $\mu$, $\rho$, and $M$. Orange dimonds correspond to constant $\mu=1$; blue circles  correspond to constant $\rho=4$; green triangles correspond to constant $M=4$. \label{fig:params}}  
\end{figure}

The analogy we draw between the pair-interaction model and colloid aggregation is important. First, it offers a mechanistic explanation for aggregation in the model, and second, the observed similarities show that the pair interaction model belongs to a broader class of models that support particle clustering through diffusive aggregation.

To conclude this section, the numerical results presented above reveal a dynamical scenario of formation and breakage of clusters in the pair-interaction model. Random encounters between particles lead to the formation of small clusters, e.g. pairs, that may subsequently grow due to further merging with other clusters. This process is balanced by the breakage of clusters due to the diffusion of individual particles. The balance between the two processes allows a steady-state regime with a patchy distribution of particle to be obtained. The pair-interaction model described a novel mechanistic framework of diffusing yet aggregating particles, both of which are crucial components for maintaining sufficiently high encounter rates between individuals in a sparce environment.


\subsection*{Dependence on the pair-interaction model parameters}

To explore the dependence of the clustering statistics on the model's parameters, we conducted a set of 34 simulations. In each simulation, we used different parameter values and calculated the clustering index averaged in time across the quasi-stationary regime. The simulations were conducted in three groups: 1) fixed $\mu$, 2) fixed $\rho$, and 3) fixed $M$. The results of the calculations are shown in Fig.~\ref{fig:params}, plotted against $\mu$, $\rho$, and $M$, revealing two characteristic dependencies. First, as anticipated above, the clustering index reaches its highest values for $\mu\ll1$ since particles have lower probabilities of escaping away from each other during their memory time. Correspondingly, as $\mu$ is increased the rate of cluster breaking increases, and the clustering index decreases. Second, for values of $M\gg1$ and $\mu\gg1$, cluster breaking still dominates formation, but in addition to that, only seldom particles are able to form pairs. Therefore, in this regime, the clustering index tends to $\langle I \rangle \rightarrow 1$, where clustering is not supported. In the case where $\mu\gg1$ but $M=1$, although clusters break frequently, particles are constantly available to make encounters, so the clustering index indicates moderate clustering formation. In contrast to that, no simple trend for the clustering dependency on $\rho$ was identified.

\subsection*{Comparison with experimental results}

\begin{figure}
	\centering
	\subfloat[]{\label{fig:S_PDF}
		\includegraphics[width=2.8in]{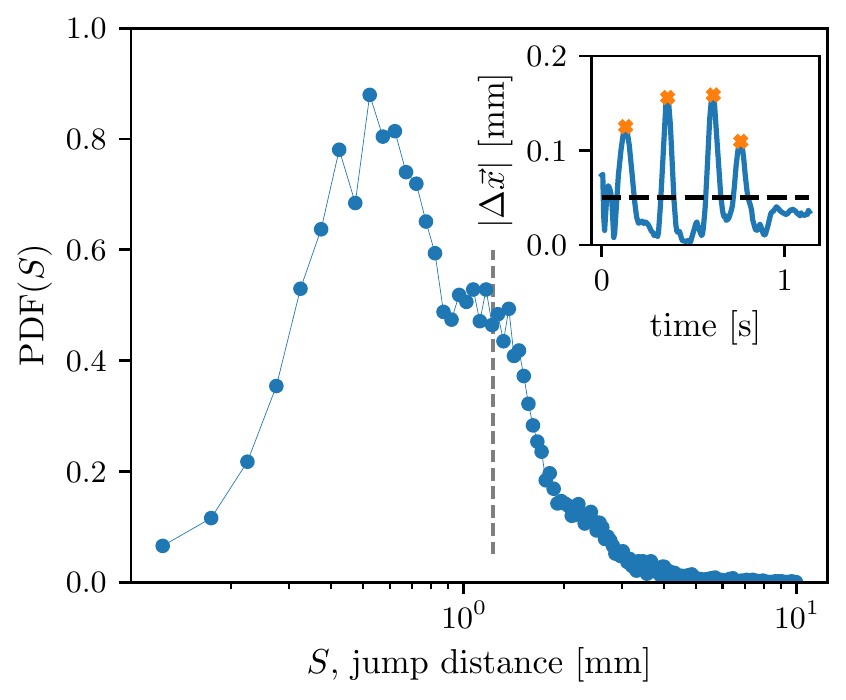}}
	\subfloat[]{\label{fig:T_PDF}
		\includegraphics[width=2.8in]{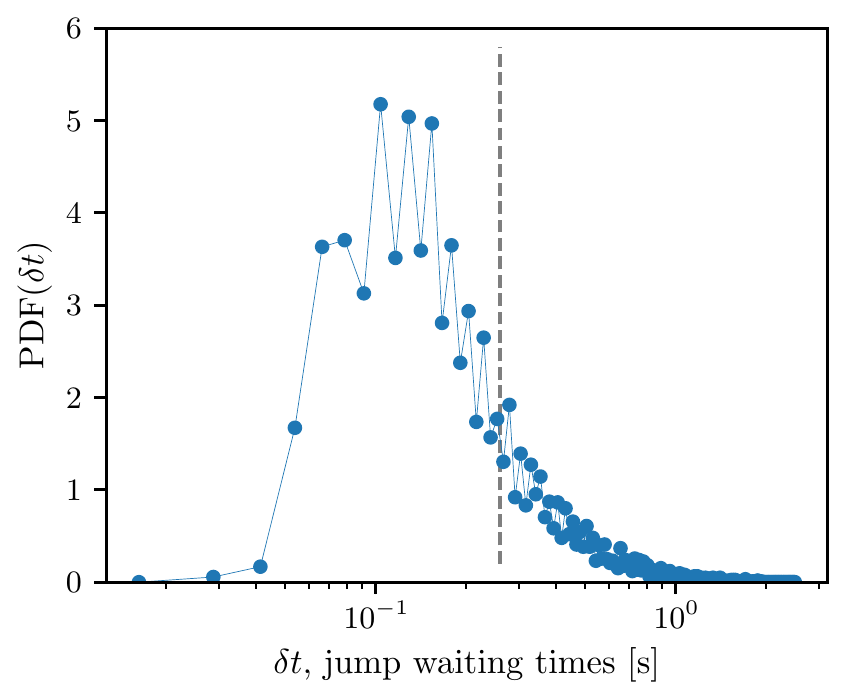}}\\
	\subfloat[]{\label{fig:NNd}
		\includegraphics[width=2.8in]{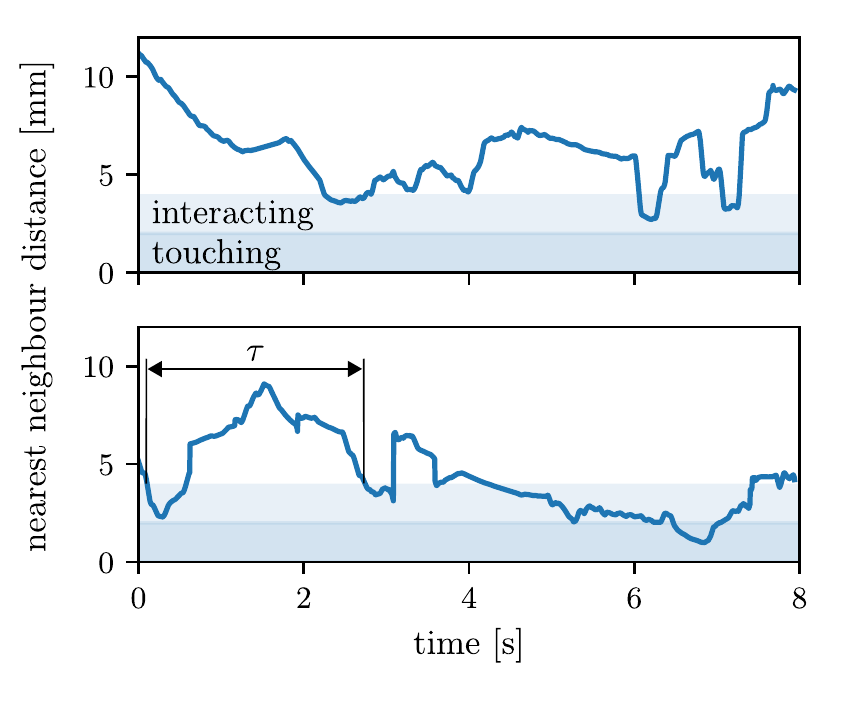}}
	\subfloat[]{\label{fig:tau_PDF}
		\includegraphics[width=2.8in]{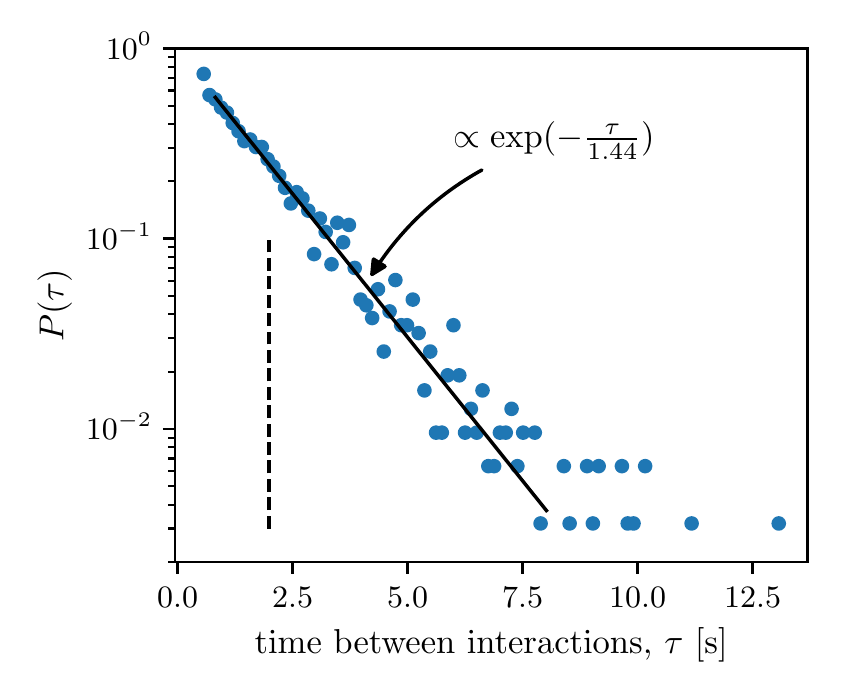}}
	\caption{(a) The inset shows the time series of displacements using time separation of $\tau=40$ ms of a copepod trajectory in the experimental record. Local peaks that were spotted by the peak searching algorithm with displacements above the 0.05 mm dashedl line threshold are marked as relocation jumps. The main pannel shows the PDF of distance traversed in relocation jumps, $S$, and the average jump length is marked by a dashed line. (b) A PDF for the watining time between two consecutive copepod relocation jumps. The average value is marked by a dashed line. (c) The distance to the nearest neighbouring copepod is plotted for two copepods as a function of time. The interaction radious, $a=4$mm, and two copepod body lengths are shown as shaded regions. The time between interactions, $\tau$, defined as the time between cosecutive events in which the nearest neighbour distance crosses the threshold $a$ from above, is demonstrated in the figure. (d) The PDF of $\tau$ for copepod trajectories in our dataset. An exponential function fit is shown as a continuous line and the average value of $\langle \tau \rangle = 1.97$ s is indicated by a vertical dashed line.}
\end{figure}

To examine the validity of the pair-interaction model to describe traits of plankton spatial distribution, we analyzed trajectories of copepods from the experiment described in the Methods section~\cite{Michalec2017, Michalec2020}. We thus start by using the recorded trajectories to estimate $\rho$ for the copepods in the experiment. Similar to particles in the model, copepods in the experiment swim by performing a series of relocation jumps. Thus, their jumping statistics could be estimated from the experimental results and used as an input for the pair-interaction model. In the inset of Fig.~\ref{fig:S_PDF}, we plot the displacement time series, defined as $\delta_\tau \vec{x} = |\vec{x}(t+\tau) - \vec{x}(t)|$, of a single copepod trajectory, where we used $\tau=25$ ms. As the figure shows, copepod jumps could be clearly identified in the time series as high local peaks of $\delta_\tau \vec{x}$. We thus used a peak searching algorithm to detect jumps in the copepod trajectories. To remove slower steady swimming periods and reject spurious peaks due to experimental noise we used a threshold of $\delta_\tau x > 0.05$ mm for detected peaks, as shown by a dashed line. Jumps that were detected by our algorithm for this example trajectory are marked in the figure. The distance traversed during each jump was calculated by comparing the positions at the beginning and end of each jump, defined by the threshold crossings of the displacement time series. The main panel of Fig.~\ref{fig:S_PDF} shows the PDF of jumping lengths calculated from the experimental data. The average jump length, $\langle S \rangle = 1.22\pm0.2$ mm, is marked with a dashed line, where the uncertainty range represents results obtained for varying $\tau$ values. In addition to that, a recent previous paper from our group~\cite{Michalec2020} used an analysis of a similar dataset of copepod trajectories, to estimate the interaction radius, where $a\approx4$ mm was estimated. This analysis gives an estimation of $\rho$ in the range 3-4 for the copepods considered here. Notably, Ref.~\cite{Kirboe2014} observed that the hydrodynamic signal of swimming copepods decays algebraically with the distance, so mechanical sensing is limited to short distances. Furthermore, the time between consecutive jumps could be estimated in cases where copepods jumped more than twice in our records. The PDF of the times passed between detected jumps is shown in Fig.~\ref{fig:T_PDF}. The average waiting time is shown as a dashed line and is equal to $\langle \delta t \rangle = 0.26$ seconds.

The second dimensionless parameter of the pair-interaction model is $\mu$ which depends on the memory time, $M$. Since no previous information is available for a memory time in copepods, we shall estimate it here for \textit{Eurytemora affinis} based on our experimental data, estimating $M$ directly from the trajectory records. To do so, we use the definition that interactions occur if two individuals are located a distance smaller than the interaction radius away from each other. Therefore, to identify interactions along swimming copepod trajectories, for each copepod trajectory we computed the distance to its nearest neighbor as a function of time. In Fig.~\ref{fig:NNd} we plot the nearest neighbor distance as a function of time for two example trajectories. For the copepod in the upper figure, there are three periods at which the distance to its nearest neighbor is smaller than the interaction radius $a=4$ mm. The copepod in the bottom figure similarly shows three periods in which the nearest neighbor distance is smaller than the interaction radius, but one of these periods is longer, lasting about 3.5 seconds and the nearest neighbor distance drops below 2mm, which suggests that the copepods are touching (since their size is about 1-1.5 mm).

Proceeding further, we can estimate the memory time, $M$, parameter by examining statistics for the time separation between the beginnings of interaction events, denoted as $\tau$. Thus, we define $\tau$ as the time that elapses between two consecutive events in which the nearest neighbor distance crosses $a$ from above, as demonstrated in the bottom panel of Fig.~\ref{fig:NNd}). In our analysis, we discarded $\tau$ samples smaller than 0.35 seconds to reject spurious oscillations due to experimental noise that would corrupt the statistics. In Fig.~\ref{fig:tau_PDF} we show PDFs of $\tau$ calculated from copepod trajectories. The PDF can be fairly approximated by an exponential function with a rate of 1.44 seconds, seen as a continuous black line. Thus, assuming that the distribution of $\tau$ is $\exp(-\tau/1.44)$ for $\tau>0.35$, we obtain an average value of $\langle \tau \rangle = 1.8$ seconds. Now, the time between interactions is expected to increase with the concentration of copepods due to the searching period, however, the high copepod concentration in our experiment suggests that individuals who are available for interactions will easily be able to find a conspecific. This consideration suggests that $\langle \tau \rangle$ will be only very slightly higher than $M$, such that $M\approx \langle \tau \rangle$ is a good approximation for our experiment. Therefore, using this approximation we obtain the estimate of $M\approx 1.8$ s, or $M/\langle \delta t \rangle \approx 6.9$. Furthermore, this memory time corresponds to the parameter value of $\mu=\sqrt{M}/\rho \approx 0.657$.

\begin{figure}
	\centering
	\includegraphics[width=2.8in]{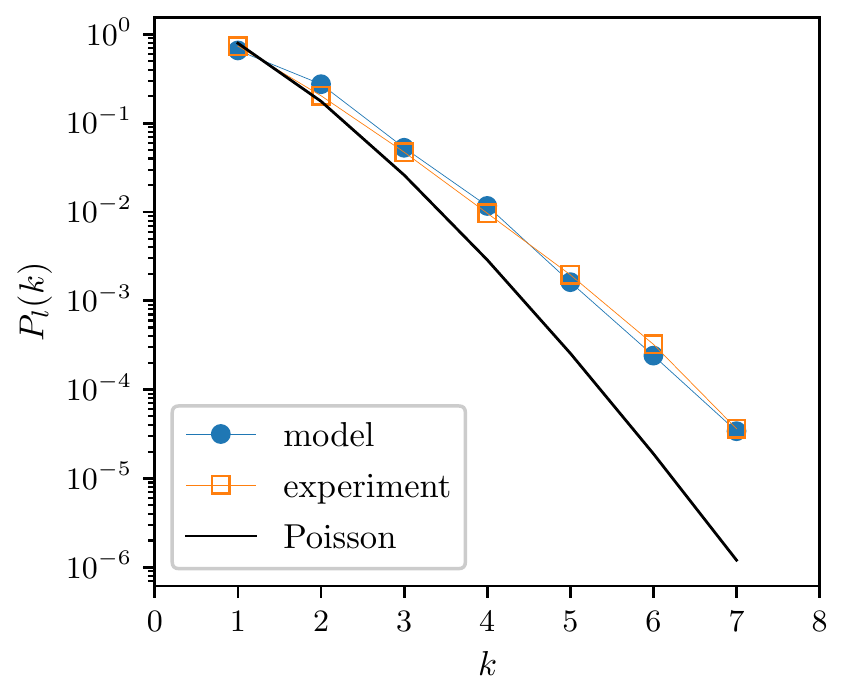}
	\caption{The PDF of cluster size  distribution for the copepods observed in the experiment, for the shuffled coordinates of the copepods in the experiment, and for a pair-interaction model simulation with $\rho=4$ and $\mu=0.66$. \label{fig:exp_PDFs}
	}
\end{figure}

To test the predictions of the pair interaction model against real copepod behaviour, we used the experimental parameters $\lambda=0.44$, $\mu=0.66$, and $\rho=4$, to run a simulation of the model. The cluster size distributions, $P_l(k)$, for this simulation is shown in Fig.~\ref{fig:exp_PDFs}. The probability for observing $k>2$ in the simulation is significantly highrer that the corresponding value of the Poisson distribution with the same $\lambda$, suggesting an increased probability for clustering ($I=1.28$). In addition to that, Fig.~\ref{fig:exp_PDFs} shows the cluster size distribution that was measured for the copepods coordinates directly, using their measured coordinates. The simulation's results and the experimental measurements are seen to agree remarkably well with each other with $r^2=0.975$. The good agreement between the experimental results and the simulation's predictions without using any fitting parameters suggests that pair interactions can lead to clustering in real copepods, which is the second key result of this work.

\section*{Conclusion}

In this work, we introduce a novel mechanistic framework that explains how interactions between pairs of individuals can support patchy particle distributions and drive the formation of mating clusters. We have thus introduced the pair-interaction model which allows exploring this scenario. Our simulations show that large clusters can be sustained in a steady state due to the balance between the pair-attraction and the diffusion that results from the individuals' active searching. The process is explained using an analogy to colloid aggregation, where high density patches form in a diffusion-limited regime and they are maintained by to the reduced diffusivity of interacting particles.

To verify the relevance of our model in describing real copepod dynamics we compared our model with the results of a laboratory 3D particle tracking experiment in still water and high animal concentration. As a first step, we obtain estimates for the model's parameters by analyzing the copepod trajectories directly. The results suggest that \textit{Eurytemora affinis} attempt to make interactions on average once about every 7 jumps. Following that, we compared the cluster size probability distribution using the directly measured copepod coordinates and the results of a simulation using the parameter values we estimated. Our results show good agreement between the experimental data and the model's results without using any fitting parameters. Therefore, our results support the hypothesis that pair-interactions can be a driver for small-scale clustering in sweeming copepods.

The animal density in our laboratory experiment was higher than in the ocean. On the one hand, this was beneficial to our analysis since it allowed us to estimate the memory time by measuring the nearest-neighbour statistics, yet on the other hand, it raises the question how closely our results apply to the typically lower animal density values observed in the ocean. Interactions in our high density experiment likely occurred mostly through hydrodynamical sensing and thus with an interaction radius of a few body sizes~\cite{Michalec2020}; however, it is known that the radius of interaction of copepods in the ocean at dilute concentration can be much larger than that, namely, orders of magnitude larger than the body size due to chemotaxis by pheromones~\cite{Yen1998,Bagoeien2005}. Thus, taking an interaction radius of $a=10$cm while keeping $S=1.2$mm and $M=1.8$s gives the model parameters $\rho\approx82$ and $\mu\approx0.03$; while $\rho$ was seen to not affect the clustering signiciantly, this low value of $\mu$ suggests a strong effect of clustering in this case (Fig.~\ref{fig:basic_characters}). Furthermore, assuming a realistic organism concentration in the ocean of 10 individuals per cubic meter~\cite{Kane2008, Seuront2001, Davis1992} gives a density of about 0.042 interaction sphere volumes per cubic meter, a value not so different from that used in the low density simulations shown in Fig.~\ref{fig:cluster_size} (the density is 0.1 for the $N_p=3000$ case), and which displayed strong clustering. These considerations suggest that our model predictions indeed apply to typical dilute concentrations and suggest that mating interactions can drive small scale clustering in the ocean.

The results we present offer a new perspective, showing that pair-interactions drive the formation of mating clusters, and thus help explain previous observations of mating aggregates in the ocean~\cite{Davis1992, PinelAlloul2007}. Furthermore, the framework we present here could be useful for understanding the role of behavior in the formation of plankton clusters at the centimeter to meter scales. Plankton clusters are likely to play important roles in biological interactions such as predation and mating, thereby influencing their life cycle and the dynamics of their populations.


\appendix

\section{Mean squared displacement for non-interacting particles} \label{App:MSD_proof}

For the case with no interactions between particles (namely, for $a=0$), the mean squared translation of particles can be calculated recursively as:
\begin{equation}
\av{|\vec{x}_i - \vec{x}_0|^2} = \av{(\vec{x}_{i-1} - \vec{x}_0)^2} + S^2 = \cdots = i\,S^2 \,\, .
\label{eq:msd}
\end{equation}
Let us prove eq.~\eqref{eq:msd} for the three-dimensional case. Without loss of generality, consider dispersion of particles starting from the origin: 
\begin{equation*}
\begin{split}
|\vec{x}_i|^2 & = x_i^2 + y_i^2 + z_i^2 \\[.5em]
& = [x_{i-1} + S\,\cos(\phi_{i-1})\,\sin(\theta_{i-1})]^2
+ [y_{i-1} + S\,\sin(\phi_{i-1})\,\sin(\theta_{i-1})]^2
+ [z_{i-1} + S\,\cos(\theta_{i-1})]^2 \\[.5em]
& = |\vec{x_{i-1}}|^2 + S^2 
+ 2S [ x_{i-1}\,\cos(\phi_{i-1})\,\sin(\theta_{i-1}) + 
y_{i-1}\,\sin(\phi_{i-1})\,\sin(\theta_{i-1}) +
z_{i-1}\,\cos(\theta_{i-1})] \,\, .\\
\end{split}
\end{equation*}
Where in the second line we used the polar representation of the translation vector, $\hat{\theta}_{i-1}\cdot S$. Taking the ensemble average of the results while recalling that the directors and the coordinate in each timestep are not correlated, we obtain the proof:
\begin{equation*}
\langle |\vec{x}_i|^2 \rangle = \langle |\vec{x}_{i-1}|^2\rangle + S^2 \,\,.
\end{equation*}

\bibliography{bib}
\bibliographystyle{unsrt}

\end{document}